\documentclass[12pt]{article}

\newcommand {\Cbar}
    {\mathord{\setlength{\unitlength}{1em}
     \begin{picture}(0.6,0.7)(-0.1,0)
        \put(-0.1,0){\rm C}
        \thicklines
        \put(0.2,0.05){\line(0,1){0.55}}
     \end {picture}}}
\newcommand {\Qbar}
    {\mathord{\setlength{\unitlength}{1em}
     \begin{picture}(0.6,0.7)(-0.1,0)
        \put(-0.1,0){\rm Q}
        \thicklines
        \put(0.2,0.05){\line(0,1){0.55}}
     \end {picture}}}
\def\IH{\relax{\rm I\kern-.18em H}}
\def\IR{\relax{\rm I\kern-.18em R}}
\def\IC{\Cbar}
\def\IQ{\Qbar}
\def\IZ{\relax{\rm Z\kern-.4em Z}}
\def\dop{{\rm d}\hskip -1pt}

\def\CN {{\cal N}}

\def\CF {{\cal F}}

\def\CH {{\cal H}}

\def\CK{{\cal K}}

\def\beq{\begin{equation}}
\def\eeq{\end{equation}}
\def\beqa{\begin{eqnarray}}
\def\eeqa{\end{eqnarray}}

\newcommand{\NP}[1]{Nucl.\ Phys.\ {\bf #1}}
\newcommand{\CQG}[1]{Class.\ Quant.\ Grav.\ {\bf #1}}
\newcommand{\PL}[1]{Phys.\ Lett.\ {\bf #1}}

\newcommand{\PR}[1]{Phys.\ Rev.\ {\bf #1}}
\newcommand{\PRL}[1]{Phys.\ Rev.\ Lett.\ {\bf #1}}

\def\one{{\hbox{ 1\kern-.8mm l}}}
\def\ii{{\rm i}}

\newcommand{\N}{{\cal N}}

\def\bb{\bar{b}}

\def\bZ{\bar{Z}}
\def\bOmega{\bar{\Omega}}
\def\bpartial{\bar{\partial}}
\def\bj{\bar{j}}

\def\L{\Lambda}
\def\bL{\bar{\Lambda}}

\newlength{\bredde}
\def\slash#1{\settowidth{\bredde}{$#1$}\ifmmode\,\raisebox{.15ex}{/}
\hspace*{-\bredde} #1\else$\,\raisebox{.15ex}{/}\hspace*{-\bredde} #1$\fi}

\begin{document}

\begin{titlepage}
\begin{flushright} KUL-TF-98/55 \\ hep-th/9812049
\end{flushright}

\vfill
\begin{center} 
{\LARGE\bf Attractors at weak gravity}
\\
\vskip 20mm  
{\bf Frederik Denef} \\
\vskip 7mm
{\em Instituut voor Theoretische Fysica\\
Universiteit Leuven \\ B-3001 Leuven, Belgium} \\
{\tt frederik.denef@fys.kuleuven.ac.be}
\end{center}
\vfill

\begin{quote}

\begin{center}
{\bf Abstract}
\end{center}
\vskip 5mm

We study the attractor mechanism in low energy effective $D=4$, ${\cal N}=2$
Yang-Mills theory weakly coupled to gravity, obtained from the effective action
of type $IIB$ string theory compactified on a Calabi-Yau manifold. Using 
special K\"{a}hler geometry, the general form of the leading gravitational 
correction is derived, and from this the attractor equations in the weak 
gravity limit. The effective Newton constant turns out to be 
spacetime-dependent due to QFT loop and nonperturbative effects. We discuss 
some properties of the attractor solutions, which are gravitationally corrected
dyons, and their relation with the BPS spectrum of quantum Yang-Mills theory.
Along the way, we obtain a satisfying description of Strominger's massless black
holes, moving at the speed of light, free of pathologies encountered in earlier
proposals.

\end{quote}

\vfill

December 1998

\end{titlepage}

\section{Introduction}

From the point of view of physics, the central issue in string theory is what
can be extracted from it at ``low'' energies in four dimensions. At least for
models with sufficient supersymmetry, this turned out to be surprisingly much,
often even {\em more} than could be found from the more conventional formulation
of four dimensional theories. This is mainly due to the beautiful, unifying
geometric picture string theory provides. In particular for four dimensional
flat space quantum field theories, many striking results have been obtained
\cite{1,2,3}. In this context, the main focus till now has been on theories
completely decoupled from gravity. However, string theory should be at its
strongest in questions {\em involving} gravity, so investigating gravitational
corrections to those results might be very fruitful, and possibly even of some
phenomenological importance.

Another area of physics in which much is expected from string theory is black
hole quantum mechanics. Interesting black hole models with still enough
supersymmetry to make their analysis manageable can be found as BPS solutions of
the low energy $D=4$, $\CN = 2$ supergravity theory obtained from $IIB$ strings
compactified on a Calabi-Yau manifold. These black holes have the remarkable
property of being {\em attractors} for the scalars in the vectormultiplets: at
the horizon, the scalars always have the same value, only determined by the
charge of the black hole, and insensitive to the variations of the scalar values
at infinity. This phenomenon was discovered by Ferrara, Kallosh and Strominger
in \cite{4} (excellent reviews can be found in \cite{5,6}), and turned out to
have a wide range of applications, also far beyond four dimensional physics. In
particular, in a fascinating paper \cite{6} (summarized in \cite{7}), Moore
uncovered a number of astonishing links between number theory and attractors.
Many of the proposals in that paper are still speculative, in part because it is
pretty hard to find tractable examples of attractors. So finding and
investigating such examples might be very fruitful, and possibly even of some
number theoretical importance.

In this paper, we try to combine both projects. We investigate the attractor
mechanism close to (but not at) a point in the moduli space of the $IIB/CY$
compactification where an effective $4D$ quantum field theory decoupled from
gravity can be isolated \cite{1,8}. The leading gravitational correction to the
field theory has a universal form, as we will show. We thus obtain attractors in
an effective quantum gauge theory weakly coupled to gravity. It turns out that
as long as their single particle mass is smaller than the Planck mass, the
attractors are no longer black holes, but QFT and gravitationally corrected
dyons, without event horizon (even for multiparticle states). Nevertheless, the
attractor property still holds, but now the attractive moduli values are reached
at the boundary of a core at finite distance from the origin. The attractive
points in moduli space are precisely the singular points where the charge under
consideration becomes massless, which could intuitively be expected from the
wrapped 3-brane picture for the BPS states. If we let the moduli at infinity
approach these attractor points and at the same time boost the solution to the
speed of light, we find a perfectly satisfactory description\footnote{The
solutions we propose are different from the related ``conifold solutions''
discussed in the literature \cite{21,21tris}, in part
due to the fact that we only insist on
continuous differentiability of a set of {\em good coordinates} on moduli space,
and not necessarily of all {\em periods} (which are not always good
coordinates).} of Strominger's massless black holes.

We furthermore find that there exist no (BPS) solutions for some charges (for
the ansatz of a spherical symmetric solution centered around a point charge).
This seems to be related to the absence of those states from the corresponding
quantum Yang-Mills BPS spectrum, though establishing a precise correspondence
will require a relaxation of the ansatz, presumably closely related to the
appearance of 3-pronged strings in the 3/7-brane picture of $\CN=2$ super 
Yang-Mills \cite{9,10,11}.

Along the way, we show that the effective Newton constant in these theories is
actually necessarily spacetime dependent. We derive an explicit formula for the
space dependence in the presence of an attractor, and find that this variation
of the Newton constant is a pure QFT loop and nonperturbative effect, in the
sense that it is absent in (electric or magnetic) weak coupling limits. This
might be of interest (as a toy model perhaps) in the light of recent claims on
measured spacetime variations of the constants of nature \cite{12}.

The structure of the paper is as follows. In section 2 we review and clarify
some basic aspects of the attractor mechanism, including multicenter solutions.
In section 3, we review the emergence of low energy Yang-Mills theories from
type $IIB$ Calabi-Yau compactifications, and derive the universal form of the
leading gravitational correction. Section 4 combines the two previous sections
to obtain the weak gravity attractor equations and their solutions. We conclude
with some comments in section 5.

An extensive review on classical dyons coupled to gravity can be found in
\cite{13}, while quantum dyons without gravity are studied e.g. in
\cite{14,15,16,17}. The principle of {\em dynamical relaxation} of BPS mass 
which underlies the
attractor mechanism, the parallel with supersymmetric Yang-Mills theory,
and quantum corrections to supersymmetric black holes in heterotic
compactifications are discussed in depth in \cite{21bis}. Black holes at
conifold points in moduli space have been considered before in
\cite{6}, \cite{21} and in particular in \cite{21tris}.

\section{Attractor mechanism}

\subsection{Invariant formalism}

We will follow the manifestly duality invariant formalism of
\cite{6}.
Consider type $IIB$ string theory compactified on a Calabi-Yau
manifold $X$. The four dimensional low energy theory is $\N = 2$ 
supergravity coupled to $n_v = h^{1,2}$ abelian vectormultiplets and
$n_h = h^{1,1} + 1$ hypermultiplets, where the $h^{i,j}$ are the Hodge
numbers of $X$. The hypermultiplets will play no role in the following
and are set to arbitrary constant values.

The vectormultiplet scalars
are identified with the complex structure moduli of $X$, and the
lattice of electric and magnetic charges with $H^3(X,\IZ)$, the
lattice of integral harmonic $3$-forms on $X$. The total
field strength $\CF$, which in this case can be identified with 
the type $IIB$ anti-self-dual five-form field strength, has values
in $\Omega^2(M_4) \otimes H^3(X,\IZ)$, where $\Omega^2(M_4)$ denotes
the space of 2-forms on the four dimensional spacetime $M_4$.
The usual components of the
field strength are obtained by picking a symplectic basis
${A^I,B_I}$ of $H^3(X,\IZ)$:
\beq
\CF = F^I B_I + G_I A^I.
\label{FGcomp}
\eeq
The total field strength satisfies the anti-self-duality constraint:
\beq
\CF = - *_{10} \CF,
\eeq
where $*_{10}$ is the Hodge star operator on the ten-dimensional
space time, which factorises according to the compactification
as $*_{10} = *_4 \otimes *_X$. This constraint relates the $F$ and
$G$ components in (\ref{FGcomp}).

The equation of motion and the Bianchi identity of the electromagnetic
field are combined in the equation
\beq
\dop \CF = 0.
\eeq
Electric and magnetic charges inside a region bounded by a surface
$S$ are found as
\beq
\Gamma_S = n_I A^I + m^I B_I = \int_S \CF
\eeq
Choosing a space/time decomposition and denoting the spatial
components of $\CF$ as $\CF_s$, the electromagnetic energy density
is simply given by:
\beq
\CH_{em} \dop t = \frac{1}{4} \int_X \CF_s \wedge *_{10} \CF_s \label{emenergy}
\eeq

\subsection{Special geometry}

The geometry of the scalar manifold of the vector multiplets,
parametrized with $n_v$ coordinates $z^a$, is special K\"ahler \cite{18}.
The metric
\beq
g_{a\bb} = \partial_a \bpartial_{\bb} \CK
\eeq 
is derived from the K\"ahler potential
\beq
\CK = - \ln ( \ii \int_X \Omega \wedge \bOmega ),
\eeq
where $\Omega$ is the holomorphic $3$-form on $X$.

Any harmonic $3$-form $\Gamma$ on $X$ can be decomposed according to 
\beq
H^3(X,\IC) = H^{3,0}(X) + H^{2,1}(X) + H^{1,2}(X) + H^{0,3}(X)
\eeq
as
\begin{equation}
\Gamma = \ii \, e^{\CK} \, \Omega \int_X \Gamma \wedge \bOmega
\; - \; \ii \, e^{\CK} \, D_a \Omega \, g^{a\bb} \int_X \Gamma \wedge 
\bar{D}_{\bb} \bOmega
\; + \; {\rm c.c.},
\end{equation}
where $D_a \Omega \equiv (\partial_a + \partial_a \CK) \Omega$.
This decomposition is useful because it diagonalizes the Hodge star
operator:
\beq
*_X \Gamma^{p,3-p}(X) = (-1)^p \, \ii \, \Gamma^{p,3-p}(X)
\eeq
This can be used to find a more explicit expression for (\ref{emenergy}). 
Denote
\beq
\eta = e^{\CK/2} \int_X \CF_s \wedge \Omega, 
\eeq
then:
\beq
\CH_{em} \dop t = \eta \wedge *_4 \bar{\eta} 
\; + \; g^{a\bb} D_a \eta \wedge *_4 \bar{D}_{\bb} \bar{\eta},
\label{emen}
\eeq
with $D_a \eta \equiv (\partial_a + \frac{1}{2} \partial_a \CK) \eta$.

\subsection{Static spherical symmetric configurations}

We take a general\footnote{This ansatz is more general than the one used
in \cite{5}. However, it reduces to the latter by the equations of
motion, as we show here.} static spherical symmetric ansatz for the metric:
\beq
ds^2 = e^{2U(r)} dt^2 - e^{-2U(r)} (\frac{1}{g(r)^2} dr^2 + r^2 d\Omega_2^2),
\eeq
or, changing variables $r = c/\sinh c \tau$, $g(r)=h(\tau)\cosh c\tau$: 
\beq
ds^2 = e^{2U} dt^2 - e^{-2U} (
\frac{1}{h^2} \frac{c^4}{\sinh^4 c\tau} d\tau^2 
+\frac{c^2}{\sinh^2 c\tau} d\Omega_2^2 \, ).
\eeq
We furthermore assume the moduli and electromagnetic fields to be spherical
symmetric and produced by a source with charge $\Gamma$ at $r=0$, that is
\beq
\CF = \omega \otimes \Gamma - *_{4} \omega \otimes *_X \Gamma,
\label{F_ansatz}
\eeq
where 
$\omega = \frac{1}{4 \pi} \sin \theta \, \dop \theta \wedge \dop \phi
= *_4(\frac{e^{2U}}{4 \pi h} \, \dop\tau \wedge \dop t)$. In the $IIB$ string theory, 
this corresponds to a three brane wrapped around 
the 3-cycle $\hat{\Gamma}$ Poincar\'{e} dual to $\Gamma$, at the origin.  
Putting all fermionic fields to zero, we find for the reduced
effective action\footnote{The action $S \sim \int F^I \wedge G_I + \cdots$
is only determined up to
symplectic duality transformations (i.e. up to choice of symplectic
basis $(A^I,B_I)$ in (\ref{FGcomp})). However, to get a manifestly 
consistent
reduced action principle at fixed field strength $\CF$, $F^I=\dop A^I$
should only appear in the action such that the action is not varied
via changes of $F^I$ when the other fields are varied. With the
ansatz we use, this is the case when we choose our basis such that
$\Gamma \cdot B_I = 0$, and then
the electromagnetic part of the action is equal to the electromagnetic
energy (\ref{emen}), leading to the above expression for $S$.}
(modulo boundary terms), in Planck units:
\beq
S = \frac{1}{2} \int_0^\infty \dop \tau \; \left\{ h(\dot{U}^2 
+ g_{a\bb} \dot{z}^a \dot{\bar{z}}^{\bb} - c^2) 
+\frac{1}{h}e^{2U} V(z)
- \frac{c^2}{\sinh^2 c\tau} (h + \frac{1}{h} - 2) \right\}, \label{toten}
\eeq
where $\dot{f} \equiv \frac{d f}{d\tau}$.
The ``scalar potential'' $V(z)$ is derived from (\ref{emen}) and 
(\ref{F_ansatz}):
\beq
V(z) = |Z|^2 + g^{a\bb} D_a Z \bar{D}_{\bb} \bZ
= |Z|^2 + 4 g^{a\bb} \partial_a |Z| \bar{\partial}_{\bb} |Z|
\eeq
with the (position dependent) ``central charge'' $Z$ defined as
\beq
Z=e^{\CK/2} \int_X \Gamma \wedge \Omega 
= e^{\CK /2} \int_{\hat{\Gamma}} \Omega.
\eeq

Now the equation of motion for $h(\tau)$ obtained from variations of (\ref{toten})
corresponding to radial diffeomorphisms 
({\em i.e.} $\delta f(\tau) = f(\tau) + \epsilon(\tau) \dot{f}(\tau)$),
acting on the fields $U$ and $z^a$ only,
actually implies $h=1 + \frac{k}{c^2} \tanh^2 c\tau$, and we can use the
1-parameter diffeomorphism freedom from the introduction of the constant $c$ to
put $k=0$, $h=1$.
Varying $h$ on the other hand implies the constraint (at $h=1$):
\beq
\dot{U}^2 + g_{a\bb} \dot{z}^a \dot{\bar{z}}^{\bb} - e^{2U} V(z) = c^2.
\label{constraint}
\eeq
However, since at $h=1$, by $\tau$-translational invariance,
the left hand side is a conserved quantity along $\tau$-translations anyway,
this just determines the value of $c$ (from the boundary conditions),
leaving no nontrivial constraint independent
of the other equations of motion. Bearing this
in mind, we can simply put $h \equiv 1$ and rewrite (\ref{toten}) as
\beq
S = \pm e^U |Z| \; \Big|_{\tau = \infty}^{\tau = 0}
+\frac{1}{2} \int_0^\infty \dop \tau \; \left\{ (\dot{U} \pm e^U |Z|)^2 
+ \|\dot{z}^a \pm 2 e^U g^{a\bb} \bpartial_{\bb}|Z| \|^2 
- c^2 \right\}.
\label{kwadr_toten}
\eeq

\subsection{BPS solutions: the attractor equations}

From (\ref{kwadr_toten}), it is clear that the reduced action (and the energy)
at fixed values of $c$ and the boundary moduli, has a minimum at
\begin{eqnarray}
\dot{U} &=& -e^U |Z| \label{at1} \\
\dot{z}^a &=& -2 e^U g^{a\bb} \bpartial_{\bb} |Z|, \label{at2}
\end{eqnarray}
(if these equations have a solution).
Equation (\ref{constraint}) then implies $c=0$, so $r=1/\tau$ and
\beq
ds^2 = e^{2U} dt^2 - e^{-2U} d\vec{x}^2.
\eeq
Assuming asymptotic flatness, {\em i.e.} $U \to 0$ at spatial infinity,
these solutions saturate the BPS bound
\beq
E = |Z(\tau=0)|.
\label{BPS}
\eeq
Here we have dropped the $\tau=\infty$ boundary term since (\ref{at1}) and
(\ref{at2}) imply that both $e^U$ and $|Z|$ are monotonously decreasing
functions (see also equation (\ref{abs_Z}) below) satisfying the estimate 
$e^U |Z| \leq \min\{|Z(0)|/(1 + |Z(\infty)| \tau), |Z| \}$, and hence  
$e^U |Z| \to 0$ when $\tau \to \infty$. Note that this estimate
also implies that when $Z(\infty) \neq 0$, the solution is a black hole
with horizon at $r=0$. A slightly more detailed analysis \cite{5}
gives for the horizon area $A=4 \pi |Z(\infty)|^2$.

Choosing the other sign possibility in (\ref{kwadr_toten}), does {not}
give an acceptable\footnote{at least not for our purposes; in \cite{19}, it is
discussed in what sense such solutions could still be meaningful.}
solution: now $e^U$ and $|Z|$ are {\em increasing}
functions, satisfying the estimate 
$e^U |Z| \geq |Z(\tau_*)|/(1-|Z(\tau_*)| \tau)$ for any fixed $\tau_*$,
hence any nontrivial solution develops a singularity at finite distance from 
the origin, and has infinite action and energy. Furthermore, these solutions
would be gravitationally repulsive and leave the weak gravity region of moduli
space. Note however that in a {\em finite
region} of spacetime, preventing $\tau$ to run to infinity, such solutions
might be acceptable and possibly important.\footnote{a comparable situation is perhaps
the occurrence of exponential ``tunnelling'' solutions of Maxwell's equations
between two dielectrics.}

Equations (\ref{at1}) and (\ref{at2}) are called the {\em attractor equations}. 
This is because their solutions converge to fixed moduli values at
$\tau = \infty$, namely to those values for which $|Z|$ is minimal. Indeed,
from (\ref{at2}), we see:
\beq
\frac{d}{dt} |Z| = -4 e^{U} g^{a\bb} \; \partial_a |Z| \, \bpartial_{\bb} |Z| < 0,
\label{abs_Z}
\eeq 
so the moduli will flow ``down the hill'' till a minimal value of $|Z|$ is
reached (with vanishing norm of its gradient).  This is intuitively
clear in the brane picture, since we expect the 3-brane to ``pull'' the
moduli such that its volume is minimized. 

The attractor equations can also be obtained from the requirement of
conservation of one half of the supersymmetries \cite{4,20,21bis,21}.
Solutions and generalisations have been discussed for example in 
\cite{20,21bis,21,21tris}. 

\subsection{Multicenter case}

The previous discussion is readily extended to the extremal multicenter
case
{\em with equal charges}\footnote{Multicenter solutions with charges
corresponding to different elements of $H^3(X,\IZ)$, and in particular
with mutually nonlocal charges (=nonzero intersection product), are
much more difficult to study, and their existence is not clear.}, by
introducing an effective ``radial'' coordinate 
\beq
\tau \equiv \frac{1}{n} \sum_{i=1}^n \tau_i,
\eeq
where $i$ runs over the $n$ different centers and $\tau_i$ is defined as
$\tau$ in the previous discussion, relative to the $i$th center.
Surfaces of equal $\tau$ can be considered as equipotential surfaces 
for the multi-source configuration. The ansatz for the metric is the
extremal
\beq
ds^2 = e^{2U(\tau)} dt^2 - e^{-2U(\tau)} d\vec{x}^2.
\eeq
The electromagnetic field is given by superposition and has exactly
the same form as (\ref{F_ansatz}), with $\omega \equiv \sum_i \omega_i
= n *_4 (\frac{e^{2U}}{4\pi} \, \dop\tau \wedge \dop t)$
(this is only the case for equal charges). The scalar fields are
supposed to be functions of $\tau$ only.

Since the complete setup is formally the same as for the spherically
symmetric case, so are the attractor equations.
Therefore, everything said about the (extremal) spherical
symmetric case applies to the 
(extremal equal charge) multicenter case as well.

\section{Weak gravity Yang-Mills limit}

Suppose we tune the moduli such that some $3$-cycles in
$X$ become very small (measured by their periods 
$Z = e^{\CK/2} \int \Omega$).
Then the corresponding BPS states (as described above) become very light
and one expects the low energy effective theory, with energies restricted
to finite values relative to the masses of those light BPS states, to be
a certain four dimensional $\CN =2$ supersymmetric quantum field theory,
with BPS spectrum given by those light charges. Since a lot is known
about the string theory low energy effective action
(including quantum corrections), this observation can be used to derive
various nontrivial results about quantum field theories. This 
``geometric engineering'' is of course well known and there exists a
very extensive literature on the subject \cite{1}. Most of those
studies focus on the rigid quantum field theory itself, completely
decoupled from gravity. However, in principle, the special geometry
setup as described above also gives gravitational corrections which should
be taken into account as long as the exact (singular) decoupling limit
is not reached. Some examples were studied in \cite{8}.
As we will show, the lowest order gravitational
correction has a universal form.

We call the regime in which the lowest order gravity corrected
theory is accurate the {\em weak gravity limit}, while  
{\em rigid limit} refers to complete decoupling from gravity.

We will focus on the case where the resulting quantum field theory
is $\CN=2$ $SU(N)$ Yang-Mills (with possibly some additional $U(1)$ factors),
though generalisations are clearly possible. To make things specific, 
let us consider an algebraic Calabi-Yau manifold $X$, in a certain
patch with affine coordinates $(x,y,w,z)$ given by a
polynomial equation $W(x,y,z,w) = 0$. The coefficients of this
polynomial are parametrized by the complex structure moduli.
Suppose we can choose a modulus $\Lambda$ such that large $\Lambda$ 
corresponds to the $SU(N)$ weak gravity limit. This can be realized
by letting $X$ degenerate locally to an $A_n$ singular $ALE$ fibration
over a sphere when $\Lambda \to \infty$ \cite{1}, 
that is, for large $\Lambda$ and small $x,y,w$:
\begin{eqnarray}
W &\approx& W_{ALE} = w^2 + y^2 + x^N + \Lambda^{-2/N} c_{N-2}(z) x^{N-2}
+ \cdots + \Lambda^{-1} c_0(z) \label{Wapprox} \\
\Omega&\approx&\frac{1}{2 \pi \ii} \,
\frac{\dop z}{z} \wedge \frac{\dop x \wedge \dop y}{w}, \label{Oapprox}
\end{eqnarray}
where the $c_i(z)$ are certain polynomial functions of $z$ and $1/z$,
dependent on a set of ``rigid'' moduli $u_i$, $i=1,\ldots,r$,
e.g. in
the case of pure $SU(N)$ Yang-Mills $c_i(z) = u_i = \mbox{const.}$
for $i \neq 0$ and $c_0(z) = \frac{1}{2}(z + \frac{1}{z}) + u_0$.

In the appendix, it is shown (under some weak extra technical
assumptions) that the universal form of
the K\"ahler potential in such a weak gravity limit is
\beq
\CK = -\ln (a \ln |\Lambda|^2 + b) 
+ \frac{|\Lambda|^{-2/N}}{a \ln |\Lambda|^2 + b} \, K(u,\bar{u}) + \cdots,
\label{Kpot}
\eeq
where $a$ and $b$ are constants {\em independent of the rigid moduli $u^i$},
and $K(u,\bar{u})$ is the Seiberg-Witten rigid special K\"ahler potential
for the limiting gauge theory. The dots include
$u$-independent terms of nonzero (positive) order in $\Lambda^{-1/N}$,
and $u$-dependent terms higher than second order in $\Lambda^{-1/N}$.

More specifically, there is a Riemann
surface $\Sigma$ fibred over the $z$-plane, given by
\beq
x^N + c_{N-2}(z) x^{N-2} + \cdots + c_0(z) = 0,
\eeq
endowed with a meromorphic 1-form
\beq
\lambda = x \frac{\dop z}{z},
\eeq
such that for any of the ``small'' Calabi-Yau 3-cycles $\Gamma$,
we have a corresponding 1-cycle $\gamma$ on the Riemann surface 
for which
\beq
Z(\Gamma)
= \frac{\Lambda^{-1/N}}{\sqrt{a \ln |\Lambda|^2 + b}} Z_0(\gamma) \label{Zwg}
\label{Zred}
\eeq
with $Z_0(\gamma) = \int_\gamma \lambda$. Then we have
\beq
K(u,\bar{u}) = Q^{ij} Z_0(\gamma_i) \bar{Z_0}(\gamma_j),
\eeq
where $\{\gamma_i\}_i$ is a basis of 1-cycles on $\Sigma$
and $Q^{ij}$ is the inverse of the intersection form 
$Q_{ij} = \gamma_i \cdot \gamma_j$.

Consistent weak gravity truncation requires the rigid moduli excitations
to be bounded and $\Lambda$ to remain sufficiently large. Then indeed
the local special geometry reduces approximately to rigid special geometry,
and the physics is well described by a lowest order gravity corrected 
quantum field theory (see also below).

\section{Attractor equations in the weak gravity limit}

\subsection{Equations}

In the following we will assume there are no other complex structure moduli 
apart from $\Lambda$ and $u^i$. That is, the low energy theory is $\CN = 2$
$SU(N)$ Yang-Mills (possibly with matter) without additional $U(1)$ factors.

The weak gravity metric on moduli space, derived from (\ref{Kpot}) is:
\begin{eqnarray}
g_{\L \bL} &\approx& 
\frac{|\L|^{-2/N}}{a\ln |\L|^2 + b} \;
\frac{a^2}{(a \ln|\L|^2 + b) |\L|^{2-2/N} }
 \\
g_{i\bj} &\approx&
\frac{|\L|^{-2/N}}{a\ln |\L|^2 + b} \;
g_{0,i\bj}
\\
g_{i\bL} &\approx&
- \frac{|\L|^{-2/N}}{a\ln |\L|^2 + b} \;
\frac{1}{\bL} \, (\frac{1}{N} + \frac{a}{a\ln|\L|^2 +b}) \; \partial_i K
\label{gil1}
\\
&\approx&
- \frac{|\L|^{-2/N}}{a\ln |\L|^2 + b} \;
\frac{1}{N \bL} \; \partial_i K.
\label{gil2}
\end{eqnarray}
where $g_{0,i\bj} \equiv \partial_i \bpartial_{\bj} K$ is the Seiberg-Witten
metric on rigid moduli space.  
For the last approximation, equation (\ref{gil2}), we have supposed that 
$\ln|\L|^2 + b/a \gg N$, which is of course satisfied when
$\Lambda \to \infty$, but which might not be the case if we
want to study large but finite $\Lambda$ (e.g. for phenomenology\footnote{For
example we consider gravity at typical accelerator energies
to be really weak, while $\ln(E_{Planck}/E_{acc})$ is only about $50$})
or the $N \to \infty$ limit.
For now, we will keep (\ref{gil1}). 

Up to a factor $\sim (1 + O(|\L|^{-2/N} \ln |\L|))$, the inverse metric is
given by
\begin{eqnarray}
g^{\L\bL} &\approx&
(a \ln|\L|^2 + b) |\L|^{2/N} \;
\frac{a \ln |\L|^2 +b}{a^2} |\L|^{2-2/N} 
\\
g^{i\bj} &\approx&
(a \ln|\L|^2 + b) |\L|^{2/N} \;
g_0^{i\bj} 
\\
g^{i\bL} &\approx&
(a \ln|\L|^2 + b) |\L|^{2/N} \;
\frac{1}{a^2} g_0^{i \bj} \bL |\L|^{-2/N} 
[\frac{1}{N}(a\ln |\L|^2 + b) + a] \, 
g_0^{i \bj} \bpartial_{\bj} K
\end{eqnarray}

We now plug this together with (\ref{Zwg})\footnote{Since we want to study the weak
gravity limit, we restrict to the small Calabi-Yau 3-cycles, which carry the
charges of the rigid quantum field theory.}
in the attractor equations
(\ref{at1}) and (\ref{at2}). For the rigid moduli derivatives, we find 
(to lowest order):
\beq
\dot{u^i} = -2 \sqrt{a \ln |\L|^2 + b} |\L|^{1/N} \; 
e^U g_0^{i\bj} \bpartial_{\bj} |Z_0|. 
\eeq
Recall that this is an expression in units 
such that the Planck mass is one. Let us change to units adapted to the
gauge theory by taking the mass of the gauge theory (BPS) 
particles to be given by 
\beq
M(\gamma) = |Z(\gamma)| M_{Pl} \equiv |Z_0(\gamma)|,
\eeq
evaluated at spatial infinity. So, from (\ref{Zwg}), we see that in
these units
\beq
M_{Pl}^2 = (a \ln|\L_\infty|^2 + b) |\L_\infty|^{2/N}. 
\eeq
We can implement this change of units by a change of coordinates 
$\tau \to \tau^\prime$ defined by
\beq
\dop \tau^\prime = \frac{1}{\sqrt{G(\tau)}} \dop\tau,
\label{repar}
\eeq
where
\beq
G \equiv \frac{|\L|^{-2/N}}{a \ln|\L|^2 + b}
\eeq
can be considered as a space-dependent Newton constant (at spatial infinity,
$G = 1/M_{Pl}^2$). When $G$ does not vary too much (which will be the case in the
weak gravity limit), this transformation is approximately just a rescaling
of $\tau$ (or $r$). Denoting derivatives with respect to $\tau^\prime$ again
with dots,
we find for the attractor equations:
\begin{eqnarray}
\dot{u^i} &=& - 2 e^U g_0^{i\bj} \, \bpartial_{\bj} |Z_0| \label{gauge} \\
\dot{U} &=& - G e^U |Z_0|  \label{gaugegrav} \\
\frac{\dot{\Lambda}}{\Lambda} &=& G e^U |Z_0| \,
(1-g_0^{i\bj} \partial_i K \bpartial_{\bj} \ln \bZ_0) \; 
(\ln |\L|^2 + b/a)^2 (\frac{1}{N} + \frac{1}{\ln |\L|^2 +b/a} ).
\end{eqnarray}
Here we see clearly that in the (singular) limit $G=0$, gravity is indeed
completely decoupled from the gauge theory degrees of freedom.

The identification of $G$ with the Newton constant can be made more
precise by rescaling the metric with a factor $1/G$ (instead of performing
the reparametrization (\ref{repar})) and check that the full bosonic 
$4D$ effective action then indeed becomes (approximately) of the form
\beq
S = \int_{M_4} d^4x \frac{1}{16 \pi G} \sqrt{g} \,R \; + \; S_{YM} + \cdots,  
\eeq
where $S_{YM}$ is the Seiberg-Witten effective action (in curved spacetime) and
the dots denote the remaining part of the action, containing e.g. the kinetic
terms for $\Lambda$.

Note also that this Newton constant identification is convention (or frame)
dependent; we choose the frame in which the Yang-Mills part of the
action (or equivalently (\ref{gauge})) becomes $\Lambda$-independent.
A discussion on the relation between different conventions with varying
constants of nature can be found in \cite{12}.

In the approximation where $\ln G$ can be considered very large as well,
we can rewrite the real part of the last equation as
\beq
\frac{d}{d\tau^\prime} (\frac{1}{(\ln G)^2 G}) \approx
e^U |Z_0| (2 - \nabla K \cdot \nabla \ln|Z_0|^2 )
\eeq

Equation (\ref{gauge}) is just what one would obtain for
BPS field configurations
from the abelian Seiberg-Witten effective action (in curved spacetime),
and equation (\ref{gaugegrav})
is the coupling of gravity to the gauge theory fields which one would
intuitively expect: it is just (the relativistic generalisation of) Newton's
law for the gravitational field of a spherically symmetric energy distribution. 
The remaining equation however, determining the variation of $G$, is more 
intricate. Notice the rigid moduli dependent factor
\beq
f \equiv 1-g_0^{i\bj} \, \partial_i K \, \bpartial_{\bj} \ln \bZ_0.
\eeq
It is straightforward to check that $f=0$ when $K$ is quadratic in the moduli and 
$Z_0$ linear. Since (for a suitable choice of coordinates in moduli space)
this is effectively the case at weak gauge coupling (electric or magnetic),
we can conclude that $f$ only differs from zero thanks to QFT loop and
nonperturbative effects, and hence that, from this point of view,
the spatial variation of $G$ is entirely a quantum effect.

\subsection{Solutions}

In the following, we drop the primes on $\tau$.
A very useful property following from (\ref{gauge}) is the constantness of the
phase of the {\em rigid} central charge:
\beq
\frac{d}{d \tau} \arg Z_0 = 0.
\eeq
Its absolute value on the other hand is a decreasing function:
\beq
\frac{d}{d\tau} |Z_0| = - e^U \|\partial Z_0\|^2.
\eeq

In the case of pure $SU(2)$ Yang-Mills at $G=0$, the attractor equations reduce
entirely to the abelian truncation\footnote{Which is obtained by sending the
free parameter $\delta$ in \cite{14} to infinity.} of the ``quantum corrected
monopole equations''
studied in \cite{14}. It is therefore not surprising that many of the features
discussed there, including the constantness of phase 
(which indeed, as anticipated in \cite{14}, extends to the multicenter case),
apply here as well. The
inclusion of the nonabelian degrees of freedom affects the exact
$\tau$-dependence of the flow in moduli space, though not really drastically.
On the other hand, since we do not include the nonabelian excitations, we don't
have an immediate obstruction to enter into the ``strong coupling'' region
of moduli space, where the electric nonabelian description becomes inadequate.

Let us study the flows in rigid moduli space in more detail.
Since $Z_0$ is analytic, its absolute value has no minima except at zeros of $Z_0$. 
Therefore, in the weak gravity limit, for our ansatz, the central charge
$Z_0$ (and $Z$) will always flow to zero.

Note that $Z_0$ can only be zero at points of marginal stability; for example in
the pure $SU(2)$ case, $0=Z_0=n a + m a_D$ implies $a_D/a = -n/m \in \IR$.
However, if the zero of $Z_0$ is at a nonsingular point of moduli space, we run
into trouble. Indeed, close to such a point, we can choose a coordinate $v$ such
that $Z_0 = v$.\footnote{$Z_0$ cannot have a double zero since this would imply
$\partial_i Z_0 = \int_{\hat{\gamma}} \partial_i \lambda = 0$ for all $i$ there,
and hence that $\hat{\gamma}$ is a vanishing cycle at this point, which therefore
would be singular.} Equations (\ref{gauge}) and (\ref{gaugegrav}) then reduce to
\begin{eqnarray}
\arg v &=& \mbox{const.} \\
\frac{d|v|}{d\tau} &=& - k e^U \\
\frac{d e^{-U}}{d\tau} &=& G |v|,
\end{eqnarray}
with $k$ a finite positive constant. So $U$ will be approximately constant, and 
\beq
|v| \approx |v_0| - k e^U (\tau - \tau_0)
\eeq
where $v_0$ is the value of $v$ at ''initial'' $\tau=\tau_0$.
At $\tau = \tau_* \equiv \tau_0 + e^{-U_0}|v_0|/k$, $Z_0=0$,
and the flow breaks down:
it is not possible to continue the BPS solution beyond this point, since
$d|v|/d\tau$ cannot be negative at $v=0$. One could try to continue
the solution by gluing it to the ``reverse BPS flow'' obtained by
changing the sign of the RHS of the attractor equations, but as
discussed above, those solutions break down as well and are not
acceptable. So we conclude that there simply are no BPS 
solutions\footnote{to the leading abelian low energy effective action,
however, since the higher derivatives of $v$ vanish when $\tau \to \tau_*$,
higher order
corrections are not likely to alter qualitatively the conclusions.}
(satisfying the ansatz) for charges for which $Z_0$ becomes zero at
a regular point of moduli space, and we expect such states to be absent
from the quantum BPS spectrum of the theory. Looking back at the
pure $SU(2)$ example, and taking into account that $-1<a_D/a<1$ on
the curve of marginal stability, we see that this happens when $|m|>|n|>0$,
states which are indeed known not to be present in the BPS spectrum.
This phenomenon is however not restricted to charges with $|m|>|n|>0$
at $\tau=0$.
Due to monodromies, charges which asymptotically do not satisfy this 
condition, can start to do so at finite $\tau$ by crossing a cut in moduli
space. 

Now if the zero of $Z_0$ occurs at a singular point of moduli space, at a
point where the cycle $\hat{\gamma}$ vanishes, as is the case e.g. for
an $SU(2)$ monopole at $u=1$, things change considerably. 
Taking again $v = Z_0$, we now have for small $v$, 
$K \sim |v|^2 \ln|v|^{-2} + \cdots$ \cite{24} and
\beq
\frac{d|v|}{d\tau} = - e^U \frac{k}{\ln |v|^{-2}} .
\eeq
Again, we reach $v=0$ at a finite $\tau=\tau_*$, but now $d|v|/d\tau$ {\em vanishes} 
at this point (as well as the first derivative of the other fields).
Therefore we can continue the solution with continuous first
derivatives to an ``interior'' ($\tau > \tau_*$) constant solution given by 
$v(\tau)=0$, $U(\tau)=U(\tau_*)$, $G(\tau) = G(\tau_*)$. 
Only the electromagnetic field strength is nontrivial in this core region
--- it is still given by (\ref{F_ansatz}) --- though it contains no energy!
This gives a physically reasonable solution (see also below), contrary to
matching to an inverted flow, which is again not acceptable.
Higher order and nonabelian
corrections could be important for the precise value of  the inverse core radius
$\tau_*$ (and could even push
it all the way to infinity), but presumably this will not alter the essential conclusions.
Indeed, in the brane picture, such a state corresponds to a 3-brane wrapped around a
vanishing cycle located at the origin, and we expect the brane to minimize it's volume
by attracting the moduli to the conifold point. 

So we conclude that the
attractors at weak gravity are precisely the conifold points in
rigid moduli space. 

This solution can be used to study Strominger's massless black holes
\cite{22}. If we
take $Z_0$ to be already zero at spatial infinity, our solution gives simply flat space
with constant moduli 
(but still the --- energyless --- electromagnetic field of a point charge).
We shouldn't expect more of course for a massless particle at rest!
However, we can let the rest mass approach zero and simultaneously boost the solution
along the $x$-axis
to approach the speed of light, keeping the energy $E = |Z_0(0)|/\sqrt{1-v^2}$
fixed.
When $\gamma \equiv 1/\sqrt{1-v^2} \to \infty$, the boosted metric is given by
\beq
ds^2 = dt^2 - dx^2 + 4 \gamma^2 U (dt-dx)^2 -dy^2 -dz^2,
\eeq
and, for nonzero $x-t$, 
$\tau \to \frac{1}{\gamma |x-t|} \equiv \frac{\chi}{\gamma}$.
Denoting $V \equiv \gamma^2 U$ and $Y_0 \equiv \gamma Z_0$, we have
$|Y_0(0)| = E$, and from the attractor equations
\begin{eqnarray}
G &=& \mbox{const.} \\
\frac{d V}{d\chi} &=& -G |Y_0| \\
\frac{d |Y_0|}{d \chi} &=& - \frac{k}{2 \ln (\gamma |Y_0|^{-1})} \to 0,
\end{eqnarray}
which implies $|Y_0| = \mbox{const.} = E$ and $V=-G E \chi$.

So we find
a simple but nontrivial solution, with $Z_0 = 0$ everywhere,
the electromagnetic field strength of a point particle boosted to the speed of light, and
a ``shockwave metric'' (due to the combined effect of expansion of the core region
in the rest frame and longitudinal Lorentz contraction while taking the limit): 
\beq
ds^2 = dt^2 - dx^2 - dy^2 - dz^2 - \frac{4 G E}{|x-t|} (dx - dt)^2.
\eeq
This is the Aichelburg-Sexl metric for a massless 
particle \cite{23}.
It is locally flat and can be brought to standard form by changing 
coordinates as
$t^\prime = t \mp 2GE \ln |t-x|$, $x^\prime = x \pm 2GE \ln |t-x|$,
where the upper (lower) sign is to be used for $t-x>0$ ($t-x<0$).

Thus we see the solutions we find are physically quite satisfactory. 
In
particular, there are no undesirable features such as
negative ADM mass,
gravitational repulsion or unphysical divergences, 
encountered in some of the earlier proposals for the description of
these states (as in \cite{21}, or the Seiberg-Witten approximation in
\cite{16}, where the solutions are essentially continued as inverted
flows beyond the attractor point).
The main reason for this is the
fact that those approaches insist on having continuous first derivatives
of {\em all} periods, while we only require continuous first derivatives of 
a set of {\em good} coordinates on moduli space. 
For example at the boundary of the core of the pure $SU(2)$
monopole, we indeed find a discontinuity for the derivative of the
``electric'' period $a$, but we consider this as an artifact of 
$a$ not being a good coordinate at the attractor point $u=1$.

Note that the above solutions never develop an event horizon as long as the
moduli at infinity stay within the weak gravity region, that is, roughly as long
as the mass of the one particle BPS state under consideration is less than the order
of the Planck mass. Considering states with $n$ particles to increase the mass
will {\em not} create a horizon, no matter how large $n$ is (at least
in this low energy approximation and for our ansatz): from the attractor
equations, it is easy to see that the solution $f_n$ for $n$ charges $\gamma$ on top of
each other can be obtained from the corresponding single charge solution $f_1$ as
\beq
f_n(\tau) = f_1(n \tau);
\eeq
in other words, the $n$ particle solutions are obtained by simply rescaling
all distances with a factor $n$. In particular, the radius of the $n$ particle
core region is $n$ times larger than the one particle core radius. As such,
there is never enough energy in a sufficiently small region of space to 
produce a black hole; the particles protect themselves from collapse by the
attractor mechanism!

However, this does not at all exclude the existence
of regular black hole solutions (with weak gravity moduli at infinity)
which are not BPS or which satisfy a different ansatz. 

Finally, in view of the above discussion, it seems reasonable to conjecture, in
the spirit of  \cite{6}, that a given charge appears in the BPS spectrum of the
(gravity corrected) quantum Yang-Mills theory under consideration, if and only if
there exists a solution of the corresponding weak gravity attractor equations.
Unfortunately, things are not that simple. Take for example pure $SU(2)$.
When one starts with a monopole solution (attracted to $u=1$), and performs
on $u$ at spatial infinity a monodromy about $u=\infty$, one encounters
at a certain point a flow containing the other singularity, at $u=-1$. If one
tries to continue the monodromy by ``pulling'' the flow across this singularity, one
finds that our ansatz does no longer yield a solution (one gets a crash-flow
with $Z_0=0$ at a regular point). On the other hand, states obtained by monodromy
from states already occurring in the BPS spectrum, without crossing a line of marginal
stability, should also be in the BPS spectrum \cite{24}.
So the conjectured correspondence
between quantum BPS states and attractor solutions can only work if we generalize
our ansatz in one way or another. This looks very much
like what happens in the type $IIB$ $3/7$ brane picture of $\CN=2$ $SU(2)$ gauge 
theory\footnote{I am grateful to R. von Unge for pointing out this connection to me.}
\cite{9,10,11}. The open strings representing the BPS states there follow
trajectories given by the attractor equations, and fail to exist exactly at the point
where our solutions fail to exist. Interpreting the strings as deformations of the
3-branes, as in \cite{17,n4}, it is clear that this is no coincidence.
In the $3/7$ brane picture, we get a 
{\em three-pronged string} instead as BPS representative. Work is in progress
to develop the corresponding picture from the field theory point of view.

\section{Conclusions}

In this paper, we studied attractors at weak gravity. We derived the form of the
leading gravitational correction to effective low energy $SU(N)$ quantum field
theories, and from this the attractor equations in the weak gravity regime. We
investigated the spatial dependence of the effective Newton constant and
discussed some properties of the (BPS) attractor solutions and their relation
with the BPS spectrum of quantum Yang-Mills theory. To establish a full
correspondence, a more general ansatz for the solutions is needed, probably
involving a combination of the ideas in \cite{11,17,n4}, presumably allowing
multicentered solutions with different charges. Having such a correspondence
would be a powerful tool to study the BPS spectrum of (gravitationally
corrected) gauge theories, with ramifications for the question of existence of
supersymmetric Calabi-Yau cycles \cite{6}. 

It would be very interesting to consider the generalization to the nonextremal
case and investigate possible phase transitions to genuine black holes, as well
as connections to the recently discussed non-BPS states in string
and Yang-Mills theory \cite{25}. It could also be interesting to explore the
phenomenological implications of the gravitational correction.

Unfortunately, the use of these solutions for the number theoretical
considerations in \cite{6} is (at first sight) limited, since they don't have
horizons and hence no macroscopic entropy to connect to counting problems.
However, the generalizations mentioned above could change this.

Finally, we want to point out that some caution is needed in interpreting these
results. Everything is done in the low energy approximation, neglecting higher
derivatives, nonabelian excitations, and the deformation of the geometry of the
Calabi-Yau and the direct product structure of spacetime due to the presence of
the brane wrapped around a CY cycle at the origin of space. It would be
worthwile to study the modifications in the picture presented here when some of
these corrections are taken into account. Some recent work relevant to this
problem can be found in \cite{16,26}.

\vskip 10mm
\noindent
{\bf \large Acknowledgements}

\vskip 5mm
\noindent
I am very grateful to G. Chalmers, G. Moore, M. Rocek and R. von Unge for useful
and stimulating correspondence, and to B. Craps, F. Roose, J. Troost, W. Troost
and A. Van Proeyen for enlightening discussions. This work was supported
by the European Commission TMR program ERBFMRX-CT96-0045. F.D. is Aspirant of
the Belgian FWO.

\appendix

\section*{Appendix}
\setcounter{section}{1}
\setcounter{equation}{0}

In this appendix, we present a general argument leading to the form
(\ref{Kpot}) of the K\"ahler potential. Explicit examples, supporting the
general results found here, can be found in \cite{8}.

We assume $X$ to be an algebraic
Calabi-Yau manifold which in the $\Lambda \to \infty$ limit
degenerates in a certain coordinate patch to the product of an infinite
cylinder (the complex plane punctured at the origin), parametrized
by the coordinate $z$, and an $A_{N-1}$ singular compact manifold $M$.
More precisely, for $\Lambda \to \infty$,
in the patch parametrized by $z,w,y,x$, 
the polynomial defining $X$ is, to order $\Lambda^{-1}$,
given by $W \approx W_{ALE} + W^\prime(w,y,x)$, where $W_{ALE}$ is 
defined as in (\ref{Wapprox}) and
$W^\prime$ is a polynomial containing terms of order $x^{N+1}$, $w^3$,
$y^3$ and higher, possibly still depending on moduli different
from the $u^i$ (which will not be important in what follows).
We furthermore assume the holomorphic $3$-form $\Omega$, to order
$\Lambda^{-1}$, to have the following natural form:
\beq
\Omega \approx \frac{1}{\pi \ii} \, \frac{\dop z}{z} \wedge 
\frac{\dop x \wedge \dop y}{\partial_w W} \label{factO}
\eeq
At large $\Lambda$, close to the locus $S: x=y=w=0$ where the singularity
develops, the polynomial $W$ describing $X$ is then given by (\ref{Wapprox}),
and $\Omega$ by (\ref{Oapprox}).

To study the $ALE$ fibration structure close to $S$, it is convenient
to rescale $w=\Lambda^{-1/2} \tilde{w}$, $y=\Lambda^{-1/2} \tilde{y}$
and $x = \Lambda^{-1/N} \tilde{x}$. ``Close to $S$'' is equivalent to
finite values of the rescaled variables, and there $X$ is approximately
described by
\beq
\tilde{w}^2 + \tilde{y}^2 + \tilde{x}^N 
+ c_{N-2}(z) \tilde{x}^{N-2}
+ \cdots + c_0(z) = 0
\eeq
while
\beq
\Omega \approx \Lambda^{-1/N} \; \frac{1}{2 \pi \ii} \,
\frac{\dop z}{z} \wedge \frac{\dop \tilde{x} \wedge \dop \tilde{y}}
{\tilde{w}}. \label{ALO}
\eeq
Now choose a basis $\{\Gamma_I\}_I$ of $3$-cycles which are compact on 
the $ALE$ fibration, i.e. finite in the rescaled variables (see e.g.
\cite{1,8} for explicit constructions).
Denote
the (nondegenerate) intersection matrix by 
$Q_{IJ} = \Gamma_I \cdot \Gamma_J$. Extend this basis with $3$-cycles
${\Gamma^\prime}_{I^\prime}$ to a basis for $H_3(X,\IQ)$, in such way
that $\Gamma_I \cdot {\Gamma^\prime}_{I^\prime} = 0$. Since $Q_{IJ}$
is nondegenerate, this is always possible (however, in general it is
{\em not} possible to construct such a basis for $H_3(X,\IZ)$). Note
that since there are no intersections of the $\Gamma$ cycles with
the $\Gamma^\prime$ cycles, there is no obstruction for
deforming the $\Gamma^\prime$ cycles away from the neighbourhood of $S$
(where the $\Gamma$ cycles are localized). We will furthermore assume that
we can keep the $\Gamma^\prime$ cycles at finite values of $x,y,w$
when $\Lambda \to \infty$ (which is not a strong assumption). 
Explicit constructions in specific examples are discussed in \cite{8}.

The $\Gamma$ periods $\int_\Gamma \Omega$ can be seen from (\ref{ALO})
to be proportional to $\Lambda^{-1/N}$. As indicated in the paper, 
equation (\ref{Zred}), they
can be reduced to Seiberg-Witten Riemann surface periods \cite{1}.

We now turn to the $\Gamma^\prime$ periods. These can be divergent
when $\Lambda \to \infty$, but since the $\Gamma$ cycles stay finite
in $x,y,w$ and away from the singularity locus $S$ 
(such that $\partial_w W$ is bounded from below), the only potential
source of divergencies is the fact that $X$ factorizes as a direct product
($W$ becomes independent of $z$) when $\Lambda \to \infty$: consequently,
some cycles can (and will) be ``stretched'' to infinity in the $z$-plane.
Since $W$ is polynomial in $z$ and $1/z$, this will give period integrals
of the form 
\beq
\int_{\Lambda^{-q}}^{\Lambda^p} \frac{\dop z}{z} \times \mbox{(finite)},
\eeq
which to leading order are proportional to $\ln \Lambda$. So we have
\beq
\int_{\Gamma^\prime_{I^\prime}} \Omega = a_I \ln \Lambda + b_I + \cdots
\eeq
where $a_I$ and $b_I$ could a priori still be dependent on all other 
moduli.
We will now show that the leading order $u$-dependent term is actually
at most proportional to $\Lambda^{-2/N} \ln \Lambda$, so that $a_I$ and
$b_I$ must be independent of the rigid moduli $u^i$.
Indeed, using (\ref{factO}), the form of $W$ and (\ref{Wapprox}), we find
\begin{eqnarray}
\frac{\partial}{\partial u^k} \int_{\Gamma^\prime} \Omega & = &
-\int_{\Gamma^\prime} \Omega \, \frac{1}{\partial_w W} 
(\partial_w
-\frac{\partial_w^2 W}{\partial_w W})
\frac{\partial W}{\partial u^k} \\
&&=\Lambda^{-1+k/N} 
\int_{\Gamma^\prime} \Omega \,
\frac{\partial_w^2 W}{(\partial_w W)^2} \, x^k.
\end{eqnarray}
Again because $\Gamma^\prime$ stays finite in $x,y,w$ and stays away
from the singular locus $S$ (such that $\partial_w W$ stays bounded 
from below), the integral factor above can be at most proportional to
$\ln \Lambda$, and since $k \le N-2$, the full period at most to
$\Lambda^{-2/N} \ln \Lambda$.\footnote{Note that this argument fails for the
$\Gamma$ periods because on these $\partial_w W$ is {\em not}
bounded from below, and indeed the $u$-derivatives of those periods
are proportional to $\Lambda^{-1/N}$.} This is what we wanted to show.

Combining all this to compute the form of the K\"ahler potential, we find
\begin{eqnarray}
{\cal K} &=& -\ln(\ii \int_X \Omega \wedge \bar{\Omega}) \label{Kdef} \\
&=& -\ln(Q^{I^\prime J^\prime} \int_{\Gamma^\prime_{I^\prime}} \Omega
\int_{\Gamma^\prime_{J^\prime}} \bOmega 
\, + \, Q^{IJ} \int_{\Gamma_{I}} \Omega
\int_{\Gamma_{J}} \bOmega ) \\
&=& -\ln(a \ln |\Lambda|^2 + b + |\Lambda|^{-2/N} K(u,\bar{u}) + \cdots ),
\end{eqnarray}
where $a$ and $b$ are $u$-independent constants and the dots include
$u$-independent terms of nonzero order in $\Lambda^{-1/N}$,
and $u$-dependent terms higher than second order in $\Lambda^{-1/N}$.
 The reason for the absence
of terms proportional to $\ln \Lambda \ln \bar{\Lambda}$ or 
$\ln(\Lambda/\bar{\Lambda})$ 
is the fact that 
$e^{-\CK} = \ii \int_X \Omega \wedge \bar{\Omega}$ must be invariant
under the monodromy $\Lambda \to e^{2 \pi \ii N} \Lambda$. Finally, expanding
the logarithm, we find
\begin{equation}
\CK \approx -\ln (a \ln |\Lambda|^2 + b) 
+ \frac{|\Lambda|^{-2/N}}{a \ln |\Lambda|^2 + b} \, K(u,\bar{u}).
\end{equation}
The presence of the divergent term can also be inferred from (\ref{factO})
and the fact that $X$ degenerates to the direct product of an infinite
cylinder and a compact manifold. The integral over the cylinder 
(parametrized by $z$) gives a logarithmically divergent factor. Note
that this divergent term would be absent for compactifications on e.g.
the direct product of a torus and $K3$ (yielding $\CN = 4$ in four 
dimensions). Such $a=0$ cases could be discussed along the same lines,
though some features will be qualitatively different.


\begin{thebibliography}{99}

\bibitem{1} S. Kachru, A. Klemm, W. Lerche, P. Mayr and C. Vafa,
{\it Nonperturbative Results on the Point Particle Limit of N=2 Heterotic
String Compactifications},
\NP{B459} (1996) 537, hep-th/9508155; \\
A. Klemm, W. Lerche, P. Mayr, C. Vafa and N. Warner,
{\it Self-Dual Strings and N=2 Supersymmetric Field Theory},
\NP{B477} (1996) 746, hep-th/9604034; \\
S. Katz, A. Klemm and C. Vafa,
{\it Geometrical Engineering of Quantum Field Theories},
\NP{B497} (1997) 173, hep-th/9609239; \\
S. Katz, P. Mayr and C. Vafa,
{\it Mirror symmetry and Exact Solution of 4D N=2 Gauge Theories I},
Adv.\ Theor.\ Math.\ Phys.\ {\bf 1} (1998) 53, hep-th/9706110; \\
A. Klemm,
{\it On the Geometry behind N=2 Supersymmetric
Effective Actions in Four Dimensions}, hep-th/9705131; \\
W. Lerche,
{\it Introduction to Seiberg-Witten Theory and its Stringy Origin}
{\it Nucl. Phys. Proc. Suppl.} 55B (1997) 83,
{\it Fortsch. Phys.} 45 (1997) 293, hep-th/9611190.

\bibitem{2} A. Hanany and E. Witten,
{\it Type IIB superstrings, BPS monopoles, and three dimensional 
gauge dynamics},
\NP{B492} (1997) 152, hep-th/9611230; \\
E. Witten
{\it Solutions of Four-dimensional Field Theories via M Theory},
\NP{B500} (1997) 3, hep-th/9703166; \\
M. Douglas and M. Li, {\it D-brane realization of $\CN = 2$ super Yang-Mills
theory in four dimensions}, hep-th/9604041; \\
P. Howe, N. Lambert and P. West,
{\it Classical M-fivebrane dynamics and quantum N=2 Yang-Mills},
\PL{B418} (1998) 85, hep-th/9710034; \\
N. Lambert and P. West,
{\it Gauge fields and M-fivebrane dynamics},
\NP{B524} (1998) 141, hep-th/9712040; \\
N. Lambert and P. West, {\it Brane dynamics and four-dimensional quantum
field theory}, hep-th/9811177.

\bibitem{3} J. Maldacena,
{\it The large N limit of superconformal field theories and supergravity},
Adv.\ Theor.\ Math.\ Phys.\ {\bf 2} (1998)231, hep-th/9711200; \\
S. Gubser, I. Klebanov and A. Polyakov,
{\it Gauge theory correlators from non-critical string theory},
\PL{B428} (1998) 105, hep-th/9802109 \\
E. Witten,
{\it Anti de Sitter space and holography},
Adv.\ Theor.\ Math.\ Phys.\ {\bf 2} (1998) 253, hep-th/9802150.

\bibitem{4} S. Ferrara, R. Kallosh and A. Strominger,
{\it N=2 Extremal Black Holes},
\PR{D52} (1995) 5412, hep-th/9508072.

\bibitem{5} S. Ferrara, G. Gibbons and R. Kallosh,
{\it Black holes and critical points in moduli space},
Nucl.\ Phys.\ {\bf B500} (1997) 75, hep-th/9702103.

\bibitem{6} G. Moore,
{\it Arithmetic and attractors},
hep-th/9807087

\bibitem{7} G. Moore,
{\it Attractors and arithmetic},
hep-th/9807056

\bibitem{8} M. Bill\'o,  F. Denef, P. Fr\`e, I. Pesando, W. Troost,
A. Van Proeyen and D. Zanon,
{\it The rigid limit in Special Kahler geometry},
\CQG{15} (1998) 2083, hep-th/9803228; \\
M. Bill\'o,  F. Denef, P. Fr\`e, I. Pesando, W. Troost,
A. Van Proeyen and D. Zanon,
{\it Special geometry of Calabi-Yau compactifications near a rigid limit},
hep-th/9801140; \\
S. Cacciatori and D. Zanon,
{\it Gravitational corrections to N=2 supersymmetric Lagrangians},
hep-th/9809202.

\bibitem{9} T. Banks, M. Douglas and N. Seiberg,
{\it Probing F-theory with branes},
\PL{B387} (1996) 278 hep-th/9605199.

\bibitem{10} A. Sen,
{\it BPS states on a three brane probe},
\PR{D55} (1997) 2501, hep-th/9608005.

\bibitem{11} M.R. Gaberdiel, T. Hauer and B. Zwiebach,
{\it Open string-String junction transitions},
\NP{B525} (1998) 117, hep-th/9801205; \\
O. Bergman and A. Fayyazuddin,
{\it String junctions and BPS states in Seiberg-Witten theory},
\NP{B531} (1998) 108, hep-th/9802033; \\
A. Mikhailov, N. Nekrasov and S. Sethi,
{\it Geometric realizations of BPS states in N=2 theories}
\NP{B531} (1998) 345, hep-th/9803142; \\
O. DeWolfe, T. Hauer, A. Iqbal, B. Zwiebach, 
{\it Constraints on the BPS spectrum of N=2, D=4 theories with A-D-E
flavor symmetry},
\NP{B534} (1998) 261, hep-th/9805220.

\bibitem{12} J. Barrow,
{\it Varying G and other constants},
gr-qc/9711084; \\
J. Webb, V. Flambaum, C. Curchill, M. Drinkwater and J. Barrow,
{\it Evidence for time variation of the fine structure constant},
astro-ph/9803165.

\bibitem{13} M. Volkov and D. Gal'tsov,
{\it Gravitating non-abelian solitons and black holes with Yang-Mills fields},
hep-th/9810070.

\bibitem{14} G. Chalmers, M. Rocek, R. von Unge, 
{\it Monopoles in quantum corrected N=2 super Yang-Mills theory},
hep-th/9612195.

\bibitem{15} J. Gauntlett, Jq. Gomis and P. Townsend,
{\it BPS bounds for world volume branes},
J. High Energy Phys. (1998) 01:003, hep-th/9711205.

\bibitem{16} N. Lambert and P. West,
{\it Monopole dynamics from the M-fivebrane},
hep-th/9811025.

\bibitem{17} C. Callan and J. Maldacena,
{\it Brane dynamics from the Born-Infeld action},
\NP{B513} (1998) 198; \\
A. Hashimoto,
{\it The shape of branes pulled by strings},
\PR{D57} (1998) 6441, hep-th/9711097.

\bibitem{18} B. de Wit, P. Lauwers and A. Van Proeyen
{\it Lagrangians of N=2 supergravity-matter systems},
Nucl. Phys. {\bf B255} (1985) 569; \\
B. Craps, F. Roose, W. Troost and A. Van Proeyen,
{\it What is special geometry?},
\NP{B503} (1997) 565, hep-th/9703082

\bibitem{19} I. Gaida, H. Hollmann and J. Stewart, {\em Classical and quantum
analysis of repulsive singularities in four dimensional extended supergravity},
hep-th/9811032.

\bibitem{20} A. Strominger,
{\it Macroscopic entropy of N=2 extremal black holes},
\PL{B383} (1996) 39, hep-th/9602111; \\
S. Ferrara and R. Kallosh
{\it Universality of supersymmetric attractors},
\PR{D54} (1996) 1525, hep-th/9603090; \\
G. Gibbons, R. Kallosh and B. Kol,
{\it Moduli, scalar charges, and the first law of black hole thermodynamics},
\PRL{77} (1996) 4992, hep-th/9607108; \\
R. Kallosh and A. Linde, {\em Black hole superpartners and fixed scalars},
\PR{D56} (1997) 3509, hep-th/9611161; \\
P. Fr\'{e},
{\it Supersymmetry and first order equations for extremal states:
monoploes, hyperinstantons, black holes and p-branes},
Nucl.\ Phys.\ Proc.\ Suppl.\ 57 (1997) 52, hep-th/9701054.

\bibitem{21bis} S.J. Rey,
{\it Classical and quantum aspects of BPS black holes in N=2, D=4 heterotic
string compactifications},
\NP{B508} (1997) 569, hep-th/9610157.

\bibitem{21} K. Behrndt, G. Lopes Cardoso, B. de Wit, R. Kallosh, D. L\"{u}st
and T. Mohaupt, {\it Classical and quantum N=2 supersymmetric black holes},
\NP{B488} (1997) 236, hep-th/9610105; \\
M. Shmakova, {\it Calabi-Yau Black Holes},
\PR{D56} (1997) 540, hep-th/9612076; \\
W. Sabra, {\it General static N=2 black holes},
Mod. Phys. Lett. {\bf A12} (1997) 2585, hep-th/9703101; \\
W. Sabra,
{\it Black holes in N=2 supergravity theories and harmonic functions},
\NP{B510} (1998) 247, hep-th/9704147; \\
K. Behrndt, D. L\"{u}st and W. Sabra,
{\it Stationary solutions of N=2 supergravity}
\NP{B510} (1998) 264, hep-th/9705169; 

\bibitem{21tris} K. Behrndt, D. L\"{u}st and W. Sabra,
{\it Moving moduli, Calabi-Yau phase transitions and massless BPS
configurations in type II superstrings},
\PL{B418} (1998) 303, hep-th/9708065.

\bibitem{22} A. Strominger,
{\it Massless black holes and conifolds in string theory},
Nucl.\ Phys.\ {\bf B451} (1995) 96, hep-th/9504090. \\

\bibitem{23} G. 't Hooft,
{\em The scattering matrix approach for the quantum black hole: an overview},
Int. J. Mod. Phys. A11 (1996) 4623-4688, gr-qc/9607022; \\
P. Aichelburg and R. Sexl,
Gen. Rel. and Gravitation 2 (1971) 303,
W. Bonner, Commun. Math. Phys. {\bf 13} (1969) 163.

\bibitem{24} N. Seiberg and E. Witten,
{\it Electric-magnetic duality, monopole condensation, and confinement in N=2
supersymmetric Yang-Mills theory},
\NP{B426} (1994) 19, hep-th/9407087.

\bibitem{25} O. Bergman,
{\it Stable non-BPS dyons in N=2 SYM},
hep-th/9811064; \\
A. Sen,
{\it D-branes on non-supersymmetric cycles},
hep-th/9812031.

\bibitem{26} R. Koch and R. Tatar,
{\it Higher derivative terms from threebranes in F theory},
hep-th/9811128.

\bibitem{n4} K. Hashimoto, H. Hata and N. Sasakura,
{\it Multi-pronged strings and BPS saturated solutions in SU(n) supersymmetric
Yang-Mills theory},
hep-th/9804164; \\
K. Lee and P. Li,
{\it Dyons in N=4 supersymmetric theories and three-pronged strings},
hep-th/9804174.

\end{thebibliography}
\end{document}